%% file: main.tex
\pdfoutput=1
\documentclass[conference]{IEEEtran}
\IEEEoverridecommandlockouts

\usepackage{cite}
\usepackage{amsmath,amssymb,amsfonts}
\usepackage{graphicx}
\usepackage{textcomp}
\usepackage{xcolor}
\usepackage{booktabs}
\usepackage[hyphens]{url}
\usepackage{hyperref}
\hypersetup{pdftitle={Who Patches the Lightning Network? Evidence from Release Lines in Gossip, 2023-2026},
pdfauthor={Saurabh Kumar}}

\input{numbers}

\begin{document}

\title{Who Patches the Lightning Network?\\
{\LARGE Evidence from Release Lines in Gossip, 2023--2026}}

\author{
\IEEEauthorblockN{Saurabh Kumar}
\IEEEauthorblockA{Bitcoin Data Labs\\
sorukumar@bitcoindatalabs.org}
}

\maketitle
\thispagestyle{plain}
\pagestyle{plain}

\begin{abstract}
\input{sections/abstract}
\end{abstract}


\input{sections/introduction}
\input{sections/background}
\input{sections/data}
\input{sections/inference}

\input{sections/dynamics}

\input{sections/platforms}
\input{sections/exposure}

\input{sections/discussion}

\section*{Acknowledgements}
The author thanks Marisha Sahay for valuable feedback and assistance with data preparation and review.

\section*{Data and Code Availability}
Code, the aggregate datasets behind every figure and table, and a script that recomputes every number quoted in the
text are available at \url{https://github.com/bitcoindatalabs/ln-upgrade-dynamics}. All stages truncate their inputs
at the data cutoff of 13~October~2026. The underlying gossip data are public; our daily gossip snapshots are
available from the author on request.

\bibliographystyle{IEEEtran}
\bibliography{references}

\end{document}

%% file: numbers.tex
\newcommand{\AtRiskClnConnectd}{186}

\newcommand{\AtRiskClnPing}{125}
\newcommand{\AtRiskClnRecent}{334}

\newcommand{\ByYearLNDAll}{76}

\newcommand{\ClnWaveDay}{18~May~2026}

\newcommand{\ClnWaveNodes}{27}
\newcommand{\ClnWaveSigned}{93}

\newcommand{\DataUsableDays}{814}

\newcommand{\ExpCapClnGossipdCutoff}{12.7}
\newcommand{\ExpCapClnGossipdDisc}{57.8}

\newcommand{\ExpCapLLBatchDisc}{6.4}

\newcommand{\ExpCapLndCloseconfDisc}{52.5}

\newcommand{\ExpClnGossipdCutoff}{24.2}
\newcommand{\ExpClnGossipdDisc}{66.6}

\newcommand{\ExpLLBatchDisc}{14.9}

\newcommand{\ExpLndCloseconfDisc}{50.6}

\newcommand{\ExpLndSweeperDisc}{27.7}

\newcommand{\FirstSeenLndMax}{176}
\newcommand{\FirstSeenLndMin}{53}

\newcommand{\FixToDiscClnGossipd}{90}
\newcommand{\FixToDiscLLBatch}{446}
\newcommand{\FixToDiscLndCloseconf}{69}

\newcommand{\FixToDiscMedian}{256}
\newcommand{\FleetCutoffNodes}{566}
\newcommand{\FleetCutoffShare}{14}
\newcommand{\FleetFirstDay}{20~June~2025}
\newcommand{\FleetPeakCap}{1.3}
\newcommand{\FleetPeakDay}{7~December~2025}
\newcommand{\FleetPeakNodes}{1{,}781}
\newcommand{\FleetPeakShare}{28}
\newcommand{\FleetPeakTotalCap}{8{,}300}
\newcommand{\FleetWaveDay}{9~December~2025}
\newcommand{\FleetWaveHours}{26}
\newcommand{\FleetWaveNodes}{327}
\newcommand{\FleetWaveSigned}{100}
\newcommand{\MedianCLNAll}{137}

\newcommand{\MedianLNDAll}{88}
\newcommand{\MedianLNDLarge}{75}
\newcommand{\MedianLNDSmall}{135}
\newcommand{\NAdvRows}{23}
\newcommand{\NAdvTotal}{30}
\newcommand{\NAnchorCLN}{117}
\newcommand{\NAnchorClient}{797}
\newcommand{\NAnchorLND}{664}

\newcommand{\NOldLndChan}{2{,}689}
\newcommand{\NOldLndLive}{434}
\newcommand{\NPlatLineChanges}{1{,}165}

\newcommand{\NResp}{12}

\newcommand{\NUpLND}{13{,}802}
\newcommand{\NVerLines}{11}
\newcommand{\NVerNodeDays}{15{,}549}
\newcommand{\NVerNodeLines}{136}
\newcommand{\NVerNodeLinesExact}{132}
\newcommand{\NVerNodes}{124}
\newcommand{\NWaveEvents}{6}

\newcommand{\NodeLineCIHi}{99}
\newcommand{\NodeLineCILo}{93}
\newcommand{\ORBehind}{0.19}
\newcommand{\ORCapacity}{1.11}
\newcommand{\ORChannels}{1.14}
\newcommand{\ORDefaultColor}{1.08}
\newcommand{\OROld}{1.01}
\newcommand{\ORTor}{1.61}
\newcommand{\OffCallBehindLeft}{11.0}
\newcommand{\OffCallBehindNodes}{310}
\newcommand{\OffCallBehindOther}{6}
\newcommand{\OffCallBehindTor}{17}
\newcommand{\OffCallMovedMax}{3}
\newcommand{\OffCallNewestDarkN}{69}
\newcommand{\OffCallNewestLeft}{23.4}
\newcommand{\OffCallNewestLeftN}{78}
\newcommand{\OffCallNewestLine}{26.04--26.06}
\newcommand{\OffCallNewestNodes}{334}
\newcommand{\OffCallNewestOther}{14}
\newcommand{\OffCallNewestTor}{32}
\newcommand{\OffCallOddsRatio}{2.5}
\newcommand{\OffCallOrdBehindMax}{9.6}
\newcommand{\OffCallOrdBehindMedian}{3.1}

\newcommand{\OffCallOrdMaxDarkShare}{23}
\newcommand{\OffCallOrdMaxDay}{5~August~2026}
\newcommand{\OffCallOrdNewestMax}{15.6}
\newcommand{\OffCallOrdNewestMedian}{3.6}
\newcommand{\OffCallOrdWindows}{21}
\newcommand{\OffCallP}{4\times10^{-5}}

\newcommand{\PPermOpClnConnectd}{0.212}
\newcommand{\PPermOpClnGossipd}{0.003}
\newcommand{\PPermOpClnPing}{0.077}
\newcommand{\PPermOpClnRecent}{0.003}

\newcommand{\PctCohortLeftClnGossipd}{32}
\newcommand{\PctCohortLeftLLBatch}{20}

\newcommand{\PctCohortStillLLBatch}{71}
\newcommand{\PctCohortStillLndSweeper}{55}

\newcommand{\PctCohortUpClnGossipd}{46}
\newcommand{\PctCohortUpLLBatch}{10}

\newcommand{\PctDownCLN}{0.8}
\newcommand{\PctDownLND}{0.7}
\newcommand{\PctIntervalNarrow}{87}
\newcommand{\PctNodeLineExact}{97}
\newcommand{\PctOldLndLive}{12.5}
\newcommand{\PctOldLndNaive}{42.3}

\newcommand{\PctPreLND}{0.1}
\newcommand{\PctStrictMultiCLN}{47}
\newcommand{\PctStrictMultiLND}{11}
\newcommand{\PctUnrecCLN}{0.1}
\newcommand{\PctUnrecLND}{1.0}
\newcommand{\PctVerLikely}{98.2}
\newcommand{\PctVerMaxOperator}{5}
\newcommand{\PctVerStrict}{99.99}
\newcommand{\PlatLineMax}{99.7}
\newcommand{\PlatLineMin}{97}

\newcommand{\PlatShareCLN}{14}
\newcommand{\PlatShareLND}{17}
\newcommand{\PlatTimeCLN}{5.4}
\newcommand{\PlatTimeLND}{4.4}
\newcommand{\PostRelLndMax}{3{,}756}
\newcommand{\PostRelLndMin}{2{,}628}
\newcommand{\PreRelLndMax}{70}
\newcommand{\PrecCLN}{98.3}
\newcommand{\PrecLND}{100.0}

\newcommand{\RRHiOpLLBatch}{3.03}

\newcommand{\RRLoOpLLBatch}{0.56}

\newcommand{\RROpClnConnectd}{1.54}
\newcommand{\RROpClnGossipd}{3.28}
\newcommand{\RROpClnPing}{1.99}
\newcommand{\RROpClnRecent}{3.87}
\newcommand{\RROpLLBatch}{1.30}

\newcommand{\RecCLN}{99.1}
\newcommand{\RecLND}{96.4}
\newcommand{\RespLndMaxRR}{1.48}
\newcommand{\RespLndMinP}{0.12}
\newcommand{\RespMaxDisp}{70}

\newcommand{\RobFleetLLBatchDisc}{12.9}

\newcommand{\RobSignedAllClnGossipdDisc}{15}

\newcommand{\RobSignedAllLLBatchDisc}{95}

\newcommand{\RobSignedLLBatchDisc}{77}

\newcommand{\TFiftyCapLndEighteen}{41}
\newcommand{\TFiftyCapLndNineteen}{67}
\newcommand{\TFiftyCapLndTwenty}{118}

\newcommand{\TFiftyClnMax}{216}
\newcommand{\TFiftyClnMin}{66}
\newcommand{\TFiftyLndMax}{79}
\newcommand{\TFiftyLndMin}{71}

\newcommand{\TFiftyLndTwenty}{79}

\newcommand{\TFiftyLndTwentyWithFleet}{45}
\newcommand{\WaveBTCPayBase}{60}
\newcommand{\WaveBTCPayDay}{8~August~2026}

\newcommand{\WaveBTCPayNodes}{114}
\newcommand{\WaveBTCPayOlder}{48}
\newcommand{\WaveBTCPaySigned}{99}
\newcommand{\WaveBTCPayTor}{26}
\newcommand{\WaveEarlyBefore}{86}
\newcommand{\WaveEarlyDay}{28~July~2025}
\newcommand{\WaveEarlyNodes}{160}
\newcommand{\WaveEarlyVersion}{0.19.2}
\newcommand{\WaveMaxDays}{67}
\newcommand{\WaveMaxNodes}{385}

\newcommand{\WaveMinDays}{19}
\newcommand{\WaveMinNodes}{114}
\newcommand{\WaveMinSignedTwo}{100}
\newcommand{\WaveNoneDay}{20~July~2024}
\newcommand{\WaveNoneNodes}{204}

\newcommand{\WaveUmbrelBaseMax}{73}
\newcommand{\WaveUmbrelBaseMin}{61}
\newcommand{\WaveUmbrelBeforeMax}{1}

\newcommand{\WaveUmbrelDaysMax}{41}
\newcommand{\WaveUmbrelDaysMin}{19}
\newcommand{\WaveUmbrelHoursMax}{27}
\newcommand{\WaveUmbrelHoursMin}{16}
\newcommand{\WaveUmbrelMaxNodes}{385}
\newcommand{\WaveUmbrelMinNodes}{262}

\newcommand{\WaveUmbrelSignedMin}{87}
\newcommand{\WaveUmbrelTorMax}{92}
\newcommand{\WaveUmbrelTorMin}{84}

\newcommand{\WkAfterAllClnGossipd}{6.2}

\newcommand{\WkAfterAllLLBatch}{1.4}

\newcommand{\WkBeforeAllClnGossipd}{1.6}

\newcommand{\WkBeforeAllLLBatch}{1.4}

%% file: sections/abstract.tex
Every implementation of Bitcoin's Lightning payment network disclosed security vulnerabilities in 2026, and every
advisory told operators to upgrade. Whether they do has never been measured, because Lightning gossip carries no software version. Matching the
feature bits a node announces against the source of every LND, Core Lightning, Eclair and LDK release, we show that
gossip reveals a node's release line after all. Where Core Lightning nodes state their version, the inferred line is
exact for \NVerNodeLinesExact{} of \NVerNodeLines{} node--line pairs. Across \DataUsableDays{} daily snapshots of the
public graph from 2023 to 2026, the network patches on release and platform schedules, not on disclosures. Half of
LND nodes upgrade within \MedianLNDAll{} days of a release; small nodes and those that skipped releases lag far
behind. The three largest upgrade waves came about a day after the Umbrel app store shipped a new LND version. Since
June~2025, \PlatShareLND\% of LND upgrades came in the 48~hours after a platform release, windows that cover
\PlatTimeLND\% of the time. Disclosures, a median \FixToDiscMedian{} days after the fix, moved few operators: of
\NResp{} we can test, only Core Lightning's of July~2026 sped up upgrades, and only among nodes one release line
behind. Its later call to restart offline was followed mostly by nodes already patched. Two months after Lightning
Labs' August~2026 batch, \PctCohortStillLLBatch\% of the LND nodes exposed at disclosure still ran an affected
release. For maintainers, the lever on the nodes that lag is the platforms that package their releases.

%% file: sections/introduction.tex
\section{Introduction}
\label{sec:intro}

The Lightning Network is a payment network built on Bitcoin. Its nodes run open-source software and hold their
operators' funds in payment channels, so a flaw in that software can put those funds at risk. Between July and
October~2026, every major Lightning implementation disclosed security vulnerabilities. Lightning
Labs published ten LND advisories on a single day~\cite{lightninglabs2026aug} and eight more a month
later~\cite{lightninglabs2026sep}; Core Lightning (CLN) asked operators who could not upgrade to restart their nodes
offline~\cite{coreln2026offline}; Eclair and LDK shipped security releases. Each advisory ends with the same advice:
upgrade. Whether operators do, and how fast, decides how long a disclosed vulnerability stays exploitable, and by
October CLN's maintainers reported attacks on unpatched nodes~\cite{coreln2026attacks}. For web servers, browsers and
Bitcoin and Ethereum full nodes, how fast operators patch has been measured for two decades
\cite{arbaugh2000windows,rescorla2003securityholes,frei2009firefox,durumeric2014heartbleed,kim2018ethereum}. For the
Lightning Network it has not, because Lightning gossip carries no software version: earlier studies could tell which
implementation a node runs~\cite{zabka2021nodeclassification,espinasa2024decisiontree,kumar2026clientdiversity}, not
which release.

We show that the release can be read from gossip after all, and use it to ask who patches the Lightning Network:
operators on their own schedule, the platforms that package the implementations, or the advisories themselves. Each
release announces a characteristic set of protocol feature bits. From the source of every release of LND, CLN, Eclair
and LDK since 2019 we derive which sets each release can produce, and match them against \DataUsableDays{} daily
snapshots of the public graph taken by our own node from May~2023 to October~2026. The answer is that the network
patches on release and platform schedules, not on disclosures. Most operators upgrade within months of a release;
the largest movements are platforms shipping a release to all their users at once; and an advisory, published long
after its fix, moves at most the operators who were already close to current. We make four contributions.

\begin{itemize}
\item \textbf{Release-line inference from gossip.} The release line of a public node can be read from its gossip.
A release line is the finest unit gossip separates; patch releases are invisible. Where a node states its version, the
inference is close to exact: on CLN nodes whose alias names the version, the most likely line is right for
\NVerNodeLinesExact{} of \NVerNodeLines{} node--line pairs. LND nodes state no version, so we check three things that
must hold if the inference is right, and all three do: line decreases are \PctDownLND\% of observed line changes, no
LND line released since 2024 was seen on more than \PreRelLndMax{} nodes before its publication, and
\PlatLineMin--\PlatLineMax\% of the nodes that changed line right after a platform shipped a new LND line landed on
the line it shipped (\S\ref{sec:inference}).
\item \textbf{Upgrade dynamics.} Most operators upgrade within months; the ones that lag are small or already behind. Half of the
LND nodes on an older line upgrade within \MedianLNDAll{} days of a release; the smallest quarter of nodes takes
\MedianLNDSmall{}, and a node already more than one line behind has about five times lower odds of moving in any
week. These figures set apart a managed fleet of up to \FleetPeakNodes{} live LND
nodes~\cite{kumar2026clientdiversity}, whose schedule would otherwise dominate them (\S\ref{sec:dynamics}).
\item \textbf{Platforms set the pace.} The largest upgrades are node platforms shipping a release to all their users
at once. Dating the releases of two platforms to the minute from public records, we find that the three largest LND
upgrade waves, \WaveUmbrelMinNodes--\WaveUmbrelMaxNodes{} nodes each, came about a day after the Umbrel app store
published a new LND version, and another followed BTCPay Server's security release. Since June~2025, the 48~hours
after such releases cover \PlatTimeLND\% of the time but hold \PlatShareLND\% of LND upgrades (\S\ref{sec:waves}).
\item \textbf{Disclosure, response and exposure.} Disclosure moved few operators, and mostly those already close to
current. For the \NAdvTotal{} LND and CLN advisories whose fix boundary gossip can see, we measure how many live nodes
and how much capacity stayed exposed; the fixes typically shipped months before disclosure (median \FixToDiscMedian{}
days). For \NResp{} disclosures we test whether exposed nodes left the affected lines faster afterwards. Only CLN's
July~2026 disclosures
raised that rate, by a factor of \RROpClnRecent{} among nodes one line behind; nodes further behind moved when
BTCPay's own advisory had their operators update. No LND disclosure produced a detectable increase: two months after
Lightning Labs' August~2026 batch, \PctCohortStillLLBatch\% of the nodes exposed at disclosure still ran an affected
release. When CLN asked operators who could not upgrade to restart offline, the nodes that went offline were mostly
already patched: \OffCallNewestLeft\% of the nodes on the newest line went offline, against \OffCallBehindLeft\% of
the nodes behind (\S\ref{sec:exposure}).
\end{itemize}

All data behind the figures and tables, and the code, are released. We report aggregates only: no list, ranking or
label of individual nodes by release, platform or exposure.

%% file: sections/background.tex
\section{Background and Related Work}
\label{sec:background}

\subsection{Gossip and feature bits}
Public Lightning nodes announce themselves and their channels through gossip~\cite{bolt7}. A node's
\texttt{node\_announcement} carries its alias, colour, network addresses and a feature vector: a bit field in which
each pair of bits marks a protocol feature as required (even bit) or optional (odd bit)~\cite{bolt9}. Implementations
set their feature vector from compiled-in defaults, which change when a release adds or retires a feature, plus
whatever options the operator switches on. Gossip carries no version string. A node re-signs its announcement when
it restarts or changes its settings, so a node that upgrades and restarts re-announces with the feature vector of
the new release.

\subsection{Node platforms}
Many public nodes are not installed by hand. Node-in-a-box distributions such as Umbrel, Start9 and RaspiBlitz,
merchant stacks such as BTCPay Server, and hosted services package an implementation and decide when their users
receive a new release. Umbrel's app store is a public git repository: a device sees a new version of its Lightning
app once the change to the app's manifest is merged~\cite{umbrel2026apps}, which dates each release to the minute.
Gossip does not name the platform a node runs on.

\subsection{Inferring implementations}
Zabka et al.\ classify public nodes into LND, c-lightning and Eclair from default channel parameters, colours and
aliases~\cite{zabka2021nodeclassification,zabka2022empirical}. Their study contains the only prior observation of an
LND upgrade we found: after LND changed its default CLTV delta from 144 to 40 in March~2019, the new value overtook
the old one around mid-December, about nine months later. They name the planning of critical software updates as a
use of implementation inference but do not measure it. Espinasa-Vilarrasa et al.\ infer the implementation from
gossip traffic or from announced feature bits with decision trees~\cite{espinasa2024decisiontree}. Kappos et al.\
note that their gossip snapshots carried no version information, and fell back on owner-reported fields for 370
nodes~\cite{kappos2021privacy}. Our previous paper used feature-bit rules to measure client concentration and the
2026 Core Lightning shutdown~\cite{kumar2026clientdiversity}. All of these stop at the implementation; none infers
which release a node runs.

Longitudinal studies of the Lightning graph cover topology, centralisation and robustness
\cite{rohrer2019discharged,seres2020topological,lin2020secondpath,martinazzi2020evolving,zabka2024centrality},
infrastructure~\cite{casas2021taproot}, privacy~\cite{romiti2021crosslayer}, channel lifecycles
\cite{grotschla2025lifecycle}, gossip convergence~\cite{gogge2023convergence} and, most recently, stability and node
lifetimes~\cite{valko2025geolocated,valko2025stability,valko2026lifetime}. None of them studies software versions.
When Lightning Labs disclosed CVE-2019-12999, they were ``surprised to find many nodes running very old versions''
but published no numbers~\cite{lightninglabs2019cve}.

\subsection{Versions in other blockchain networks}
Bitcoin and Ethereum peers report their software version in the handshake, so crawlers can count versions directly.
Kim et al.\ found 68\% of Geth nodes two or more releases behind and tracked client age against release
dates~\cite{kim2018ethereum}. Neudecker reports that Bitcoin nodes adopt a release mostly in its first two months,
and documents an alternative client that went from zero to more than 700 peers in a
day~\cite{neudecker2019characterization}; Park et al.\ and Eisenbarth et al.\ survey Bitcoin node versions and their
known vulnerabilities~\cite{park2019bitcoinnodes,eisenbarth2021bitcoin}. SRLabs estimated that about 40\% of Parity
nodes were still vulnerable a month after an advisory~\cite{srlabs2019patchgap}. Zcash sidesteps slow upgrades by
expiring old releases at network upgrades~\cite{daniel2019mapz}. The Lightning Network offers neither a version
string to count nor a forced expiry.

\subsection{Patch deployment}
The window of vulnerability, from disclosure to patch, was framed by Arbaugh et al.~\cite{arbaugh2000windows}.
Rescorla probed servers affected by the OpenSSL vulnerabilities of 2002 and found more than two thirds still
vulnerable two weeks after disclosure~\cite{rescorla2003securityholes}. Frei et al.\ showed that automatic
updates dominate how fast a client population patches~\cite{frei2009firefox}, and Sarabi et al.\ that the vendor's
patch speed matters more than users' promptness~\cite{sarabi2017patchme}. Durumeric et al.\ followed Heartbleed
across the Internet and found that direct notification raised patching by about half~\cite{durumeric2014heartbleed};
Li et al.\ found that most notified operators never act~\cite{li2016notifications}. Nappa et al.\ modelled patch
deployment with survival analysis~\cite{nappa2015clones}, and Li and Paxson measured the life cycle of security fixes
from the code side~\cite{li2017patches}. We bring this lens to a network where versions must first be inferred, and
where a release is often fixed months or years before the vulnerability is disclosed.

%% file: sections/data.tex
\section{Data and Vantage Point}
\label{sec:data}

\paragraph{Gossip snapshots}
We take a daily dump of the public graph (\texttt{describegraph}) from our own LND node and keep, per day, every
node announcement (feature bits, alias, colour, addresses, last-update time) and every channel direction (the policy
a node sets for forwarding towards its peer, with capacity, the \emph{disabled} flag and last-update time). An
announcement's last-update time is the moment its node signed it, so the first announcement that shows a new release's
bits dates an upgrade more finely than the daily snapshot does (\S\ref{sec:waves}). The snapshots run from May~2023
to the data cutoff of 13~October~2026. A per-day quality check discards dumps that
are partial (fewer than half the channels of their period), stale (fewer than half the channels updated within two
weeks) or error responses written while the node's wallet was locked; \DataUsableDays{} days remain. Daily coverage is
dense from November~2023, with gaps in May--June~2024 (partial dumps), August~2024 and January--May~2025 (wallet
locked after restarts). Upgrade times observed across a gap are known only to lie inside it; we treat them as
interval-censored (\S\ref{sec:dynamics}).

\paragraph{Vantage}
All snapshots are one node's view. LND prunes channels that have not been updated for about two weeks, and our
view widened in July~2024 (from about 8.4k to 11--13k nodes with channels), after the node's own peer set grew. We
compare levels within a period and report shares, which are less sensitive to the view; in particular, the few
live nodes on releases from before 2021 appear only from July~2024. Only announced (public) nodes are visible:
mobile wallets and services behind unannounced channels, where much of LDK and Eclair runs, are not.

\paragraph{Who is still there}
Nodes that stop running do not leave the graph at once: their peers mark the channels towards them disabled, and
the channels stay announced until they are closed or pruned. An offline node cannot announce anything, so liveness
must be read from its peers' side~\cite{kumar2026clientdiversity}. We call a node \emph{live} on a day when at least
one channel direction towards it, set by a peer, is enabled and was updated in the previous 14~days; its \emph{live
capacity} is the capacity of those channels. Results with 7- and 28-day windows are in
Table~\ref{tab:robustness}. The distinction matters. On the cutoff day, \NOldLndChan{} LND nodes with channels
(outside the managed fleet described next) announced a release older than 0.19, but only \NOldLndLive{} of them were
live; counting every announced node would put the share of LND nodes on those releases at \PctOldLndNaive\%, against
\PctOldLndLive\% of live nodes.

\paragraph{A managed fleet}
Our previous paper found a managed fleet of small LND nodes with aliases of the form \texttt{LNT.<name>}, one or two
channels each, mostly to a handful of hubs with the same prefix, and counted up to 1,980 announced fleet
nodes~\cite{kumar2026clientdiversity}. We use its definition (the alias prefix, or at least half of a node's channels
to a fleet hub) but count live nodes. The fleet appeared on \FleetFirstDay{} and peaked on \FleetPeakDay{} at
\FleetPeakNodes{} live nodes, \FleetPeakShare\% of all live LND nodes, with \FleetPeakCap{}~BTC of live capacity
against \FleetPeakTotalCap{}~BTC for live LND as a whole; at the cutoff \FleetCutoffNodes{} of its nodes were still
live (\FleetCutoffShare\% of live LND nodes). Its nodes share one Lightning Terminal feature set and upgrade in step
(\S\ref{sec:waves}), so they are not independent operators, and we exclude them from every per-node statistic.

%% file: sections/inference.tex
\section{Inferring Release Lines from Gossip}
\label{sec:inference}

\subsection{Release lines}
We built a table of the \texttt{node\_announcement} feature bits of every release of the four implementations from
their source at each release tag: LND 0.7--0.21.4, Core Lightning (CLN) 0.8.2--26.06.9, Eclair 0.3--0.14.3 and LDK
0.0.113--0.2.7. For each release we record the bits it sets by default, the bits an operator can add through
options, flags, plugins or known custom extensions, and the default bits an operator can switch off; each entry
cites the source file it comes from. A release's date is its publication date (GitHub release page; crates.io for
LDK), the day operators could install it. Consecutive releases with identical defaults form a \emph{release line}, the
finest unit gossip can resolve: LND has 10 lines from 0.8 to 0.21, CLN 13, Eclair 13 and LDK 6. Some lines are
single minor releases (LND 0.18, 0.19, 0.20, 0.21); others span several (LND 0.15.1--0.17; LDK 0.0.120--0.1 covers
January~2024 to December~2025). Patch releases never change the defaults, so a patch level is never visible.

For a node with feature set $B$, a line $\ell$ with defaults $D_\ell$, optional bits $O_\ell$ and removable bits
$R_\ell$ is \emph{consistent} when $D_\ell \setminus R_\ell \subseteq B \subseteq D_\ell \cup O_\ell$. The
\emph{strict range} is every consistent line released by the day of observation; it contains the running line under
any non-developer configuration. The \emph{likely line} is the consistent line that needs the fewest non-default
bits among the lines released by the day the node's current feature set first appeared. Dating the inference to
that day matters: a node cannot run a line that did not exist when it began announcing its bits, and without it a
new release would relabel every node whose existing bits happen to fit the new line (233 CLN nodes running 24.05 with
the onion-message option would have ``upgraded'' to 24.08 on its release day). A set consistent only with lines not
yet released is a pre-release build; a set consistent with no line is unrecognised.

\subsection{Implementation}
Release lines are defined per implementation, so the implementation comes first. We use the client inference of our
previous paper~\cite{kumar2026clientdiversity}: implementation-specific feature bits decide most nodes, operators
whose implementation is publicly known are attributed directly, and weak signals (LND's default colour or alias,
default CLTV deltas) label the rest with low confidence. The release table adds one step. When a set is consistent
with the lines of exactly one implementation, that implementation is assigned, which recovers releases older than the
distinguishing bits, such as LND 0.8--0.11 and CLN before 23.05; and when a bit rule names an implementation that has
no release producing the set while exactly one other implementation does, the table wins (the bits unique to LND and
to Eclair are never overridden). Only operator-attributed and high-confidence labels receive a release line.
Announcements that are not visible on a day (last update~0) carry the node's previous feature set forward.

\subsection{Validation}
\label{sec:validation}
\paragraph{Implementation}
\NAnchorClient{} nodes name their implementation in their alias (``\ldots{}[LND]'', ``\ldots{}-CLN'', ``eclair'',
``ldk-node''); LND's default alias is excluded because the rules use it. On the latest day each was announced with
channels, the labels reach precision \PrecLND\% and recall \RecLND\% for LND (\NAnchorLND{} nodes) and precision
\PrecCLN\% and recall \RecCLN\% for CLN (\NAnchorCLN{} nodes) (Table~\ref{tab:validation}). The misses are almost
all low-confidence labels, which receive no release line.

\begin{figure}[t]
\centering
\includegraphics[width=\columnwidth]{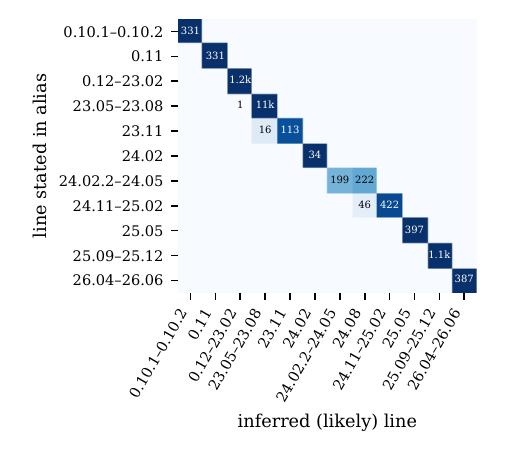}
\caption{Release line stated in CLN developer aliases (rows) against the likely line inferred from feature bits
(columns), in node-days. Shading is the row share.}
\label{fig:validation}
\end{figure}

\paragraph{Release line}
CLN built in developer mode puts the version into the default alias (e.g.\ \texttt{YELLOWWATER-v25.09}); the alias
follows the binary, so it is correct on every day the node is announced. Over \NVerNodeDays{} announced node-days of
\NVerNodes{} such nodes on \NVerLines{} lines (no operator above \PctVerMaxOperator\% of node-days), the strict range
contains the stated line on \PctVerStrict\% of node-days and the likely line is exact on \PctVerLikely\%
(Fig.~\ref{fig:validation}). Node-days of one node are not independent, so the node is the better unit: of the
\NVerNodeLines{} combinations of node and stated line, \NVerNodeLinesExact{} are exact on every day
(\PctNodeLineExact\%, 95\% CI \NodeLineCILo--\NodeLineCIHi\%). The misses are a 24.05 node with the onion-message
option, which makes it identical to 24.08; a 25.02.1 node and a 23.11.1 node whose switched-off options make them look
one line older; and a single day of a release candidate.

\paragraph{Checks without ground truth}
No LND node in our data states its version in its alias, so for LND, where most of our dynamics are measured, we rely
on checks that a wrong release table would fail. First, operators rarely return to an older release, while a noisy
inference would make nodes flicker between lines: among consecutive observations of live nodes, the likely line
decreases in \PctDownLND\% of line changes for LND (against \NUpLND{} increases) and in \PctDownCLN\% for CLN. Second,
a line assigned the wrong bits would show nodes on it long before its release: each LND line released since 2024 was
seen on at most \PreRelLndMax{} nodes before its publication, against \PostRelLndMin--\PostRelLndMax{} in its first
90~days. Third, a check from outside gossip: a platform that ships a new LND version moves its nodes to a known line
(\S\ref{sec:waves}). Of the \NPlatLineChanges{} line changes whose new announcement was signed within 48~hours after
one of the four platform releases that introduced a new LND line, \PlatLineMin--\PlatLineMax\% landed
on the line that release shipped.

\paragraph{What gossip cannot resolve}
Among live nodes with a release line, the strict range spans more than one line on \PctStrictMultiLND\% of LND and
\PctStrictMultiCLN\% of CLN node-days in 2026, because options let an older line imitate a newer one; the likely line
is unique on all of them. Unrecognised sets are \PctUnrecLND\% (LND) and \PctUnrecCLN\% (CLN) of node-days,
and pre-release builds about \PctPreLND\% of LND node-days. Three limits remain. Patch releases are invisible.
A node that upgrades without restarting, or whose new announcement does not reach us, keeps its old bits. Developer
overrides and forks can imitate any set. The first two bias exposure upward, the third in either direction; none
can be measured from gossip alone, which is why we report the strict and likely ranges side by side.

\input{tables/table_validation}

%% file: tables/table_validation.tex
\begin{table}[t]
\centering\small
\caption{Validation against self-identifying aliases. Top: client labels of 797 nodes whose alias names their implementation. Bottom: release lines of CLN nodes whose alias states the running version, graded on every announced node-day.}
\label{tab:validation}
\begin{tabular}{lrrr}
\toprule
Client & Nodes & Precision (\%) & Recall (\%) \\
\midrule
LND & 664 & 100.0 & 96.4 \\
CLN & 117 & 98.3 & 99.1 \\
Eclair & 9 & 58.3 & 77.8 \\
LDK & 7 & 100.0 & 100.0 \\
\midrule
\multicolumn{3}{l}{CLN nodes / lines with a stated version} & 124 / 11 \\
\multicolumn{3}{l}{Node-days graded} & 15,549 \\
\multicolumn{3}{l}{Strict range contains the stated line (\%)} & 99.99 \\
\multicolumn{3}{l}{Likely line equals the stated line (\%)} & 98.2 \\
\multicolumn{3}{l}{Node--line pairs exact on every day} & 132 / 136 \\
\multicolumn{3}{l}{\quad 95\% CI (\%)} & [93, 99] \\
\bottomrule
\end{tabular}
\end{table}

%% file: sections/dynamics.tex
\section{How Lightning Nodes Upgrade}
\label{sec:dynamics}

\subsection{Adoption of new release lines}
\begin{figure*}[t]
\centering
\includegraphics[width=\textwidth]{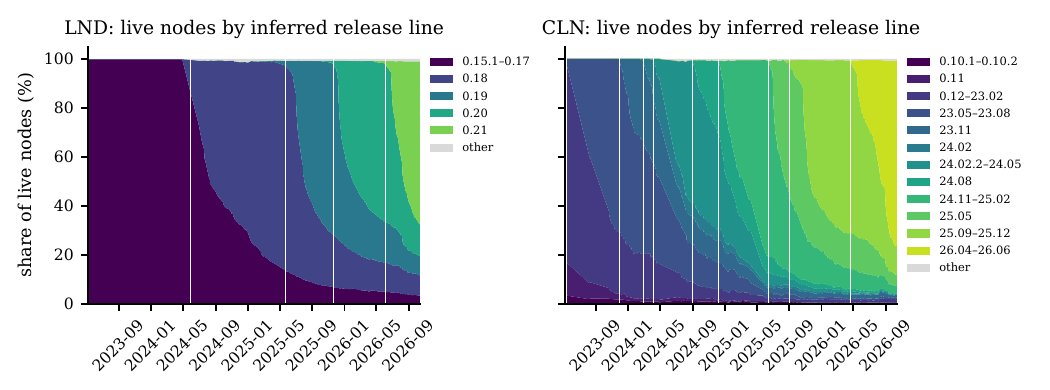}
\caption{Live LND and CLN nodes by inferred release line (likely line), excluding the managed fleet. White lines mark
releases. Lines under 2\% of nodes throughout are grouped as ``other''.}
\label{fig:linemix}
\end{figure*}

\begin{figure*}[t]
\centering
\includegraphics[width=\textwidth]{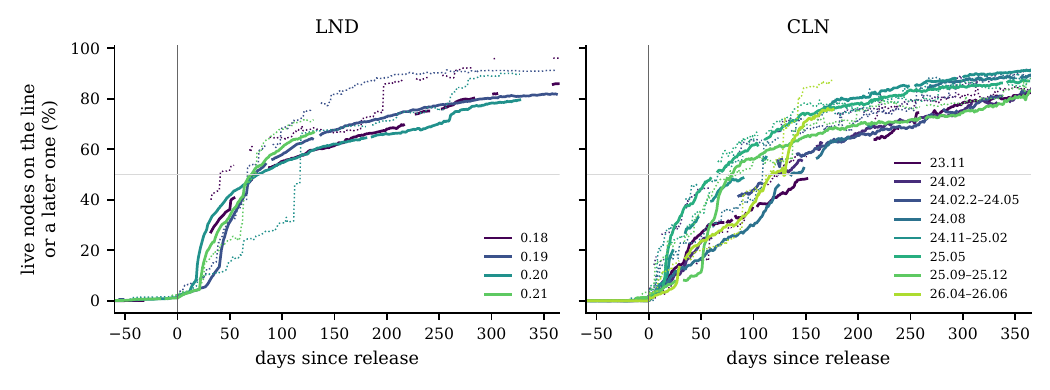}
\caption{Share of live nodes (solid) and of their live capacity (dotted) on each release line or a later one, by
days since the line's release. Lines are broken across snapshot gaps longer than three days.}
\label{fig:adoption}
\end{figure*}

Fig.~\ref{fig:linemix} shows the release-line mix of live LND and CLN nodes. Each new LND line displaces its
predecessor within a few months but leaves a tail: at the cutoff, \PctOldLndLive\% of live LND nodes were still on a
line older than 0.19, released in May~2025. New lines reached half of live LND nodes \TFiftyLndMin--\TFiftyLndMax{}
days after release (Fig.~\ref{fig:adoption}); half of live capacity moved as fast or faster for 0.18 and 0.19
(\TFiftyCapLndEighteen{} and \TFiftyCapLndNineteen{} days) but more slowly for 0.20 (\TFiftyCapLndTwenty{} days).
Counting the fleet as independent nodes would put LND 0.20 at 50\% of nodes after \TFiftyLndTwentyWithFleet{} days
instead of \TFiftyLndTwenty{}. CLN lines took \TFiftyClnMin--\TFiftyClnMax{} days. Every LND line from 0.18 on was
announced by a few nodes \FirstSeenLndMin--\FirstSeenLndMax{} days before its release: developers and early adopters
running unreleased builds on mainnet.

\subsection{Time to upgrade}
\begin{figure*}[t]
\centering
\includegraphics[width=\textwidth]{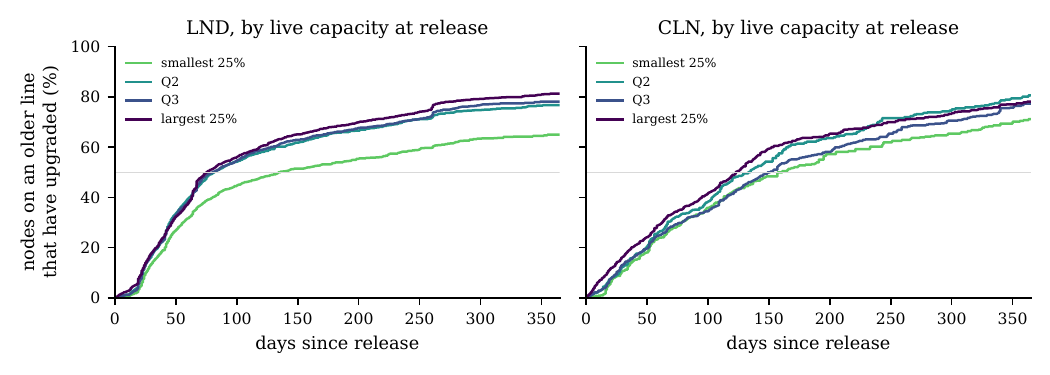}
\caption{Share of nodes that were live on an older line at a release and have since been seen on the new line or a
later one, by quartile of live capacity at the release (Turnbull estimate; LND 0.19--0.21, CLN 23.11--26.04).}
\label{fig:ttu}
\end{figure*}

Adoption curves mix upgrades with nodes arriving and leaving. To follow operators, we take every node that is live
on an older line on the release day of LND 0.19, 0.20 or 0.21 or of a CLN line from 23.11 on, and measure the time
until it is first seen on the new line or a later one. Snapshot gaps leave the upgrade day inside an interval (for
\PctIntervalNarrow\% of LND upgrades it is at most two days wide); nodes that stop being live before upgrading are
censored at their last live day. The share upgraded by day $t$ is the nonparametric maximum-likelihood estimate for
interval-censored data~\cite{turnbull1976empirical}. Half of the LND nodes upgraded within \MedianLNDAll{} days and
\ByYearLNDAll\% within a year; for CLN the median is \MedianCLNAll{} days. Size matters at the bottom: the smallest
quarter of LND nodes by live capacity took a median \MedianLNDSmall{} days, the largest quarter \MedianLNDLarge{}
(Fig.~\ref{fig:ttu}).

Table~\ref{tab:hazard} fits a discrete-time model of the weekly probability of upgrading for LND. Larger nodes
upgrade sooner (odds ratio \ORCapacity{} per tenfold capacity, \ORChannels{} per tenfold channel count), and so do
Tor-only nodes (\ORTor{}), which make up most of the platform waves of \S\ref{sec:waves}. A node more than one line
behind when the release ships is far less likely to move (\ORBehind{}): the tail is made of nodes that skipped releases before.
Nodes with LND's default colour are no slower (\ORDefaultColor{}), and nodes seen before 2024 upgrade at the same
rate as newer ones (\OROld{}).

\input{tables/table_hazard}

%% file: tables/table_hazard.tex
\begin{table}[t]
\centering\small
\caption{What predicts upgrading: discrete-time logit on node-weeks after each LND release (week and release fixed effects; standard errors clustered by node). Odds ratios with 95\% CI.}
\label{tab:hazard}
\begin{tabular}{lrr}
\toprule
Covariate & Odds ratio & 95\% CI \\
\midrule
Capacity ($\times$10) & 1.11 & [1.07, 1.16] \\
Channels ($\times$10) & 1.14 & [1.06, 1.24] \\
Tor-only & 1.61 & [1.52, 1.71] \\
LND default colour & 1.08 & [1.03, 1.14] \\
Seen before 2024 & 1.01 & [0.96, 1.07] \\
$>$1 line behind & 0.19 & [0.18, 0.21] \\
\midrule
Nodes / node-weeks & \multicolumn{2}{r}{7,239 / 220,525} \\
\bottomrule
\end{tabular}
\end{table}

%% file: sections/platforms.tex
\section{Platforms Set the Pace}
\label{sec:waves}

\subsection{Dating platform releases}
Gossip does not name the platform a node runs on, but the signing time of a node's first announcement on a new line
can be set against platform releases dated to the minute. For Umbrel we take the merge time of every pull request
that changed the version of its LND or CLN app from 2024 to the cutoff~\cite{umbrel2026apps}. A BTCPay Server
deployment updates only when its operator runs the update, which installs the LND or CLN version BTCPay pins at that
moment, so its routine version bumps reach nodes over weeks; we take its security release 2.4.2 of 7~August~2026,
which asked every operator to update at once~\cite{btcpay2026advisory}. Start9, RaspiBlitz, myNode and hosted
services publish through channels we cannot date to the day and are not covered.

\subsection{Upgrade waves}
\begin{figure*}[t]
\centering
\includegraphics[width=\textwidth]{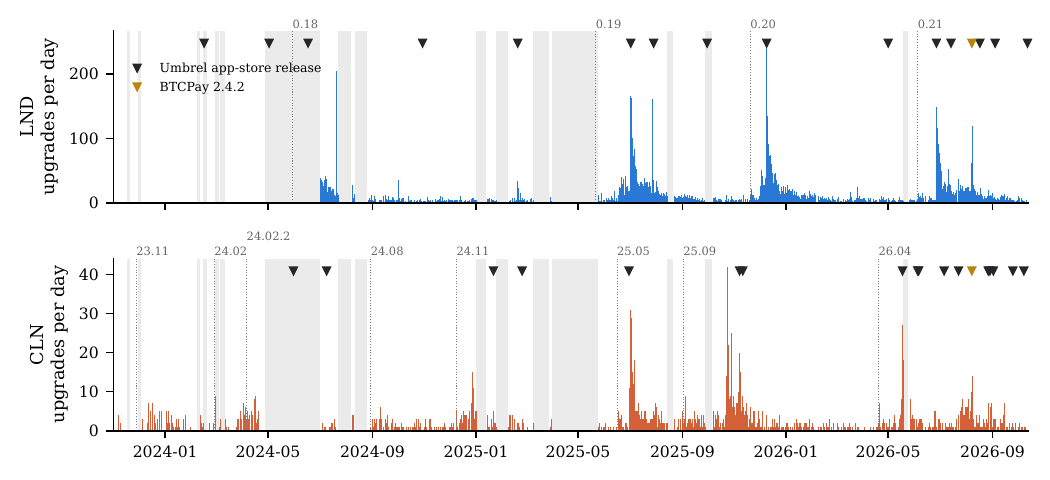}
\caption{Nodes first seen on a newer release line per day, where the upgrade day is known to within two days
(managed fleet excluded). Grey: snapshot gaps; dotted lines: LND and CLN releases; triangles: Umbrel app-store
releases (black) and BTCPay Server's security release 2.4.2 (gold) of that implementation.}
\label{fig:waves}
\end{figure*}

Upgrades arrive in bursts (Fig.~\ref{fig:waves}). We call a day a wave day for a line when more nodes moved to it
than on 99\% of all (day, line) pairs, and merge consecutive wave days on one line into an event. Outside the fleet,
LND had \NWaveEvents{} events of \WaveMinNodes--\WaveMaxNodes{} nodes. None came at the line's release; all came
\WaveMinDays--\WaveMaxDays{} days later. They are real upgrades, not gossip catching up after a gap: in every
event, \WaveMinSignedTwo\% of the nodes had signed their new announcement within two days before we first saw it.

The three largest were set by Umbrel, not by LND. Each followed Umbrel's release of a new LND version (0.19.1, 0.20.0
and 0.21.1), which came \WaveUmbrelDaysMin--\WaveUmbrelDaysMax{} days after LND's own. The timing is Umbrel's: in
each wave, at least \WaveUmbrelSignedMin\% of the \WaveUmbrelMinNodes--\WaveUmbrelMaxNodes{} nodes signed their new
announcement within 48~hours after Umbrel merged the release (median \WaveUmbrelHoursMin--\WaveUmbrelHoursMax{}~hours
after), and at most \WaveUmbrelBeforeMax\% in the 24~hours before. These nodes are far more often Tor-only than the
rest of the network (\WaveUmbrelTorMin--\WaveUmbrelTorMax\%, against \WaveUmbrelBaseMin--\WaveUmbrelBaseMax\% of all
live LND nodes). Even the managed fleet of \S\ref{sec:data} kept Umbrel's schedule: \FleetWaveNodes{} of its nodes
moved to LND 0.20 on \FleetWaveDay{}, \FleetWaveSigned\% of them within 48~hours after Umbrel published 0.20.0, with
the same median lag (\FleetWaveHours{}~hours) as the platform's other users.

A fourth event followed BTCPay Server's security release: \WaveBTCPayNodes{} nodes on \WaveBTCPayDay{},
\WaveBTCPaySigned\% of them signed within 48~hours after it. Unlike the Umbrel waves, these nodes are mostly clearnet
(\WaveBTCPayTor\% Tor-only, against \WaveBTCPayBase\%), and \WaveBTCPayOlder{} of them came from lines older than
0.20, skipping at least one release. They look like operators who had not moved to a newer LND line in months and
did so when the advisory for their merchant software told them to update; that advisory concerned BTCPay itself, not
LND. Of the remaining two events, one (\WaveEarlyNodes{} nodes, \WaveEarlyDay) was signed before Umbrel merged LND
\WaveEarlyVersion{} (\WaveEarlyBefore\% in the 24~hours before), so that release cannot explain it, and one
(\WaveNoneNodes{} nodes, \WaveNoneDay) matches no release we can date. CLN shows the same pattern at its smaller
scale: its one attributed event, \ClnWaveNodes{} nodes on \ClnWaveDay{}, followed Umbrel's release of CLN 26.04.1,
\ClnWaveSigned\% of them within 48~hours.

\subsection{Beyond the waves}
Platform releases account for a share of upgrades out of proportion to the time they cover. Since June~2025,
\PlatShareLND\% of LND upgrades whose day is known to within two days were signed within 48~hours after a platform
release that ships the new line, although these windows cover \PlatTimeLND\% of the time; for CLN, \PlatShareCLN\%
against \PlatTimeCLN\%. Since the platforms we cannot date are left out, these shares are lower bounds. The pushes
also check the inference: the nodes that changed line right after a platform shipped a new LND line landed on the
line it shipped (\S\ref{sec:validation}).

%% file: sections/exposure.tex
\section{Did Disclosure Move Operators?}
\label{sec:exposure}

\subsection{Advisories gossip can see}
We collected every public security advisory for LND, CLN, Eclair and LDK from January~2022 to the cutoff (release
notes, Lightning Labs' advisory index, Delving Bitcoin and authors' write-ups), recording the affected versions, the
first fixed release and the disclosure date, each from the source cited in Table~\ref{tab:advisories}. An advisory is
visible in gossip when its fix boundary coincides with the start of a release line (LINE), or falls inside a line but
leaves whole lines affected or fixed on either side (PARTIAL, which gives bounds). Fixes inside the newest line, such
as CLN 26.06.7--26.06.9 and Eclair 0.14.1--0.14.3, are patch releases and invisible. Eclair and LDK have too few live
public nodes with a known line (about ten Eclair nodes) for shares to be meaningful, so Table~\ref{tab:advisories}
lists the \NAdvRows{} LND and CLN entries, covering \NAdvTotal{} advisories.

Two clocks run for each advisory. Fixes usually ship long before the vulnerability is disclosed: the median gap
between the first fixed release and disclosure is \FixToDiscMedian{} days, and the six advisories Lightning Labs
published on 11~August~2026 concerned bugs fixed in LND 0.19.0, released \FixToDiscLLBatch{} days earlier. A node
still on an affected line at disclosure has therefore usually skipped a release that was already available for many
months. Exposure is the share of live nodes (excluding the managed fleet) on an affected line, with the share of
their live capacity alongside. It is largest where the fix is youngest: when Teinturier disclosed that LND accepted
too few confirmations on cooperative closes, \FixToDiscLndCloseconf{} days after the fix shipped in 0.21.0,
\ExpLndCloseconfDisc\% of live LND nodes, holding \ExpCapLndCloseconfDisc\% of live LND capacity, were on an
affected line.

\input{tables/table_advisories}

\subsection{Response to disclosure}
\label{sec:response}
If operators act on advisories, nodes should leave the affected lines faster after a disclosure than before it. For
every advisory with whole lines affected and enough snapshots on both sides (at least six of the eight weeks before
and three of the four weeks after), we take, at the start of each week, the live nodes on a wholly affected line,
and count those first seen on a later line within the week. The rate is moves per node-week, and we compare the
weeks before and after disclosure by their rate ratio. Moves are not independent: a platform push or a group of
operators moves many nodes in the same week, and the weekly counts vary up to \RespMaxDisp{} times more than Poisson
counts would. The 95\% confidence intervals are therefore quasi-Poisson, scaled by that dispersion, and we add an exact
permutation test of whether the weekly rates after disclosure exceed those before. Platform pushes could produce a
jump that no operator chose, so we also report the ratio without moves signed within 48~hours after a platform
release that ships the new line (\S\ref{sec:waves}). The comparison has a known bias: the rate of leaving a line
falls as the line ages and its most attentive operators leave, so the before-window has the higher baseline, most of
all when a disclosure follows a release by only a few months. A ratio near one therefore rules out only a large
response.

\begin{figure*}[t]
\centering
\includegraphics[width=\textwidth]{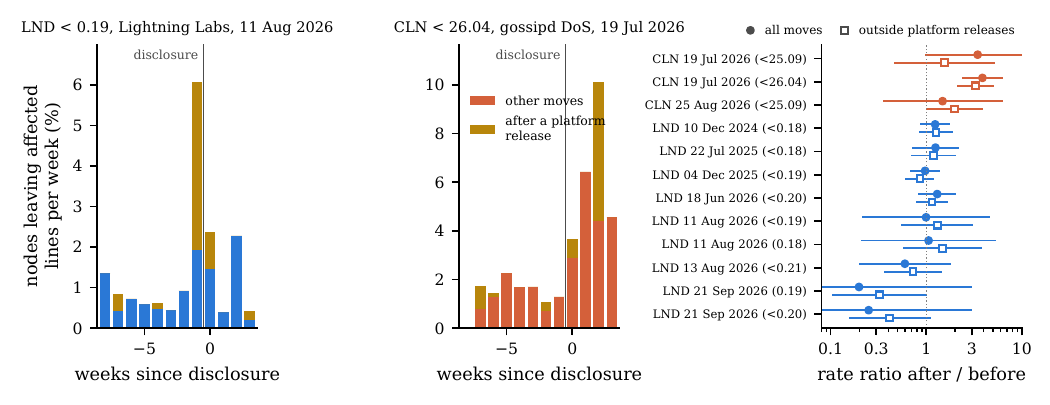}
\caption{Left and centre: weekly share of the live nodes on an affected line that moved to a later line, around two
disclosures; gold: moves signed within 48~hours after a platform release that ships the new line. Right: rate ratio
after/before disclosure with quasi-Poisson 95\% CI for every disclosure that can be tested, for all moves (filled) and outside
platform releases (open). Managed fleet excluded.}
\label{fig:response}
\end{figure*}

Twelve disclosures qualify, nine for LND and three for CLN (Table~\ref{tab:response},
Fig.~\ref{fig:response}). Only CLN's two disclosures of 19~July~2026, memory-exhaustion vulnerabilities in gossipd
(fixed in 26.04) and connectd (fixed in 25.09)~\cite{pratap2026clndos}, show a clear increase. Among live CLN nodes on
lines before 26.04, the weekly rate of leaving them rose from \WkBeforeAllClnGossipd\% before disclosure to
\WkAfterAllClnGossipd\% after. Without the moves that followed platform releases the ratio is \RROpClnGossipd{}
(permutation $p=\PPermOpClnGossipd$), the only one of the twelve tests below a Bonferroni threshold of $0.05/12$.

The operators who responded were those already close to current. The two fix boundaries split the affected nodes
into the \AtRiskClnRecent{} one line behind, on 25.09--25.12, and the \AtRiskClnConnectd{} on older lines. We compare
these two groups because, once we dated BTCPay's release, it explained most moves from the older lines; the
comparison was therefore made after seeing the data. The nodes one line behind moved steadily (indented row of
Table~\ref{tab:response}): without platform moves their rate rose \RROpClnRecent-fold, and each of the four weeks
after disclosure had a higher rate than any of the seven before (permutation $p=\PPermOpClnRecent$, the smallest
value the test can give with eleven weeks). The nodes on older lines moved mostly in the 48~hours after BTCPay's
security release of 7~August, which installed CLN 26.06.1 on every BTCPay deployment that updated; without those
moves their ratio is \RROpClnConnectd{} ($p=\PPermOpClnConnectd$). They moved when their platform's own advisory had
their operators update, not when CLN disclosed.

A month later, the ping-flood vulnerability, fixed at the same 25.09 boundary, was
disclosed~\cite{cestari2026pingflood}, and the next day the maintainers asked operators to upgrade or restart
offline~\cite{coreln2026offline}. The \AtRiskClnPing{} nodes still on older lines may have moved somewhat faster, but
not clearly: outside platform releases, which takes BTCPay's push out of the before-window, the ratio is
\RROpClnPing{} ($p=\PPermOpClnPing$). These rates also miss part of the response: a node that restarted offline as
asked left the live set and counts as not having moved. \S\ref{sec:fates} measures who left.

No LND disclosure produced a detectable increase. Across the nine, the ratios are at most \RespLndMaxRR{}, and
none is significant (permutation $p\geq\RespLndMinP$, with or without platform moves). The intervals are wide,
however. For Lightning Labs' batch of 11~August~2026, the weekly rate was \WkBeforeAllLLBatch\% before and
\WkAfterAllLLBatch\% after; without the moves that followed BTCPay's security release four days before, the ratio is
\RROpLLBatch{} (\RRLoOpLLBatch--\RRHiOpLLBatch), compatible with a modest increase but not with one as large as
CLN's (\RROpClnRecent). The exposed cohort settles what the rates leave open (\S\ref{sec:fates}): two months after
the batch, only \PctCohortUpLLBatch\% of the nodes exposed at disclosure had upgraded. The lowest ratios belong to the cooperative-close disclosure, \FixToDiscLndCloseconf{} days after 0.21.0,
and the two of 21~September~2026, which followed the platform pushes of June and August: the bias described above.

\input{tables/table_response}

\subsection{What became of the exposed nodes}
\label{sec:fates}
\begin{figure*}[t]
\centering
\includegraphics[width=\textwidth]{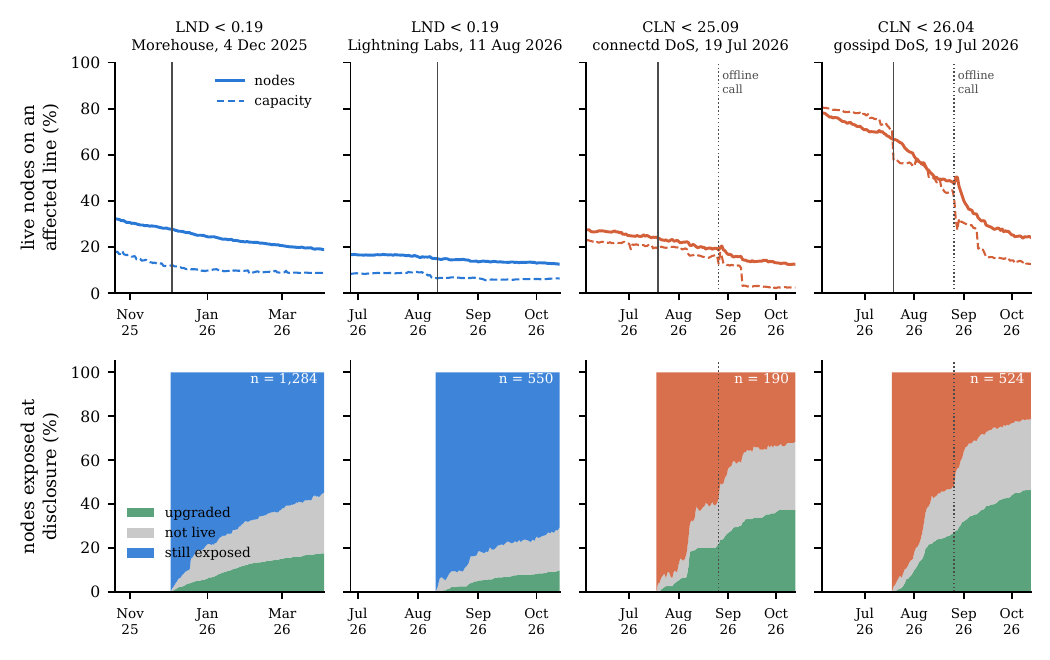}
\caption{Top: share of live nodes (solid) and of their live capacity (dashed) on a line affected by the advisory;
solid vertical line: disclosure; dotted: CLN's call of 26~August~2026 to upgrade or restart offline. Bottom: the nodes
exposed on the last snapshot before disclosure, by what became of them (still exposed in the implementation's
colour). Managed fleet excluded.}
\label{fig:cases}
\end{figure*}

Most LND nodes exposed at disclosure stayed exposed. On the day Lightning Labs disclosed its batch,
\ExpLLBatchDisc\% of live LND nodes, holding \ExpCapLLBatchDisc\% of live LND capacity, were on a line older than 0.19
(Fig.~\ref{fig:cases}). By the cutoff two months later, \PctCohortStillLLBatch\% of them were still live on an
affected line; \PctCohortUpLLBatch\% had upgraded and \PctCohortLeftLLBatch\% were no longer live. Morehouse's
disclosure of three LND vulnerabilities fixed in the same release, on
4~December~2025~\cite{morehouse2025stalling,morehouse2025failback2,morehouse2025inbox}, found \ExpLndSweeperDisc\% of live LND nodes
exposed; four months later \PctCohortStillLndSweeper\% of that cohort was still live on an affected line. The exposed
LND nodes are small: their share of capacity is about half their share of nodes.

CLN's exposure fell much further: from \ExpClnGossipdDisc\% of live CLN nodes at the gossipd disclosure to
\ExpClnGossipdCutoff\% at the cutoff (from \ExpCapClnGossipdDisc\% to \ExpCapClnGossipdCutoff\% of capacity). Of the
exposed nodes, \PctCohortUpClnGossipd\% had upgraded and \PctCohortLeftClnGossipd\% were no longer live. Much of the
fall is nodes leaving rather than upgrading, as in the CLN shutdown we measured
earlier~\cite{kumar2026clientdiversity}; leaving also ends the exposure, and restarting offline was what the
maintainers asked of operators who could not upgrade.

Release lines show who followed that call: mostly operators already patched against both July vulnerabilities. Of
the \OffCallNewestNodes{} live CLN nodes on the newest line (\OffCallNewestLine) on 25~August, \OffCallNewestLeft\%
were no longer live two days later, against \OffCallBehindLeft\% of the \OffCallBehindNodes{} on older lines (odds
ratio \OffCallOddsRatio, Fisher's exact test $p=\OffCallP$). Both are well above the medians of the
\OffCallOrdWindows{} two-day windows of August before the call (\OffCallOrdNewestMedian\% and
\OffCallOrdBehindMedian\%). The highest of those, \OffCallOrdNewestMax\% and \OffCallOrdBehindMax\% from
\OffCallOrdMaxDay, came during the closing waves of early August~\cite{kumar2026clientdiversity}, when only
\OffCallOrdMaxDarkShare\% of the nodes that left still had announced channels; at the call, \OffCallNewestDarkN{} of
the \OffCallNewestLeftN{} newest-line nodes that left still had theirs, so they had gone offline rather than lost
their channels. Transport does not explain the gap: among Tor-only nodes \OffCallNewestTor\% and \OffCallBehindTor\%
left, among the others \OffCallNewestOther\% and \OffCallBehindOther\%. Nor did the nodes behind upgrade instead: at
most \OffCallMovedMax\% of either group changed line in those two days, and the next CLN release came out on
28~August. The nodes still exposed to the July vulnerabilities mostly neither upgraded nor left, which is also why
CLN's exposure briefly rose at the call (Fig.~\ref{fig:cases}).

\subsection{Robustness}
\label{sec:robust}
Table~\ref{tab:robustness} varies the definitions behind the four case studies. A 7- or 28-day liveness window moves
the exposure by about one percentage point. Counting the fleet as nodes would lower LND's exposure, since the fleet
ran fixed lines, to \RobFleetLLBatchDisc\% instead of \ExpLLBatchDisc\% for Lightning Labs' batch. The strict lower
bounds for LND sit well below the likely-line exposure, because a newer line with some default features switched off
can produce the bits of an older one (\S\ref{sec:inference}); the likely line assumes the fewest such changes. The
exposed nodes are not stale entries in our view of the graph: \RobSignedLLBatchDisc\% of the LND nodes exposed to
Lightning Labs' batch had signed their current announcement, with the affected line's feature bits, in the 14~days
before disclosure (against \RobSignedAllLLBatchDisc\% of all live LND nodes). CLN nodes re-sign far less often
(\RobSignedAllClnGossipdDisc\% of all live CLN nodes within 14~days), so CLN exposure rests more on the absence of a
new announcement.

\input{tables/table_robustness}

%% file: tables/table_advisories.tex
\begin{table*}[t]
\centering\footnotesize
\caption{Advisories whose fix boundary is visible in gossip (LND and CLN). Exposure = share of live nodes of that implementation whose inferred release line is affected, in \%; share of their live capacity in parentheses; for PARTIAL rows (the fix lands inside a line) the bounds are given in brackets at disclosure. Fix$\to$disc.\ = days from the first fixed release to public disclosure. \# = advisories grouped in the row.}
\label{tab:advisories}
\resizebox{\textwidth}{!}{%
\begin{tabular}{lllrlrrrr}
\toprule
Impl. & Advisory (disclosed) & Class & \# & Fixed in & Fix$\to$disc. & At disclosure & +60\,d & Cutoff \\
\midrule
CLN & Fake channel DoS~\cite{morehouse2023fakechannel} (2023-08-23) & PARTIAL & 1 & 23.02 & 175 & -- & 6 (2) & 1 (0) \\
CLN & Replacement cycling (CVE-2023-40234)~\cite{riard2023replacement} (2023-10-16) & PARTIAL & 1 & 23.08.1 & 34 & 38 [24--100] (17) & 26 (10) & 2 (0) \\
CLN & BOLT11 parsing bugs, batch 1~\cite{morehouse2023clninvoice} (2023-12-08) & PARTIAL & 1 & 23.02 & 282 & 4 [4--95] (2) & 2 (1) & 1 (0) \\
CLN & BOLT11 parsing bugs, batch 2~\cite{morehouse2023clninvoice} (2023-12-08) & LINE & 1 & 23.11 & 10 & 96 (96) & 69 (73) & 3 (0) \\
CLN & funding\_created assertion DoS~\cite{pratap2026clnassert} (2026-05-16) & LINE & 1 & 25.09 & 256 & 29 (24) & 24 (21) & 12 (2) \\
CLN & connectd queue memory exhaustion~\cite{pratap2026clndos} (2026-07-19) & LINE & 1 & 25.09 & 320 & 24 (20) & 14 (3) & 12 (2) \\
CLN & gossipd unknown-SCID memory exhaustion~\cite{pratap2026clndos} (2026-07-19) & LINE & 1 & 26.04 & 90 & 67 (58) & 29 (16) & 24 (13) \\
CLN & Ping-flood OOM~\cite{cestari2026pingflood} (2026-08-25) & LINE & 1 & 25.09 & 357 & 19 (16) & -- & 12 (2) \\
LND & Fake channel DoS~\cite{morehouse2023fakechannel} (2023-08-23) & PARTIAL & 1 & 0.16.0 & 147 & -- & 0 (0)$^\dagger$ & 1 (0) \\
LND & Replacement cycling (CVE-2023-40233)~\cite{riard2023replacement} (2023-10-16) & PARTIAL & 1 & 0.17.0 & 13 & 0 [0--100] (0)$^\dagger$ & 0 (0)$^\dagger$ & 1 (0) \\
LND & Premature channel\_update OOM~\cite{delving2023lndgossip} (2023-12-21) & LINE & 1 & 0.14.0 & 764 & 0 (0)$^\dagger$ & 0 (0)$^\dagger$ & 1 (0) \\
LND & channel\_update signature after rate limit~\cite{delving2023lndgossip} (2023-12-21) & PARTIAL & 1 & 0.15.0 & 546 & 0 [0--0] (0)$^\dagger$ & 0 (0)$^\dagger$ & 1 (0) \\
LND & Onion bomb (CVE-2024-38359)~\cite{morehouse2024onionbomb} (2024-06-18) & PARTIAL & 1 & 0.17.0 & 259 & -- & -- & 1 (0) \\
LND & Irrevocable fees / update\_fee breach fee-burn~\cite{harding2024irrevocable} (2024-12-10) & PARTIAL & 1 & 0.18.3 & 89 & 33 [33--100] (25) & 24 (11) & 4 (0) \\
LND & Excessive HTLC failback~\cite{morehouse2025failback} (2025-03-04) & LINE & 1 & 0.18.0 & 278 & 20 (8) & -- & 4 (0) \\
LND & gossip\_timestamp\_filter goroutine DoS~\cite{morehouse2025gtsfilter} (2025-07-22) & PARTIAL & 1 & 0.18.3 & 313 & 11 [11--60] (3) & 9 (2) & 4 (0) \\
LND & Replacement stalling, failback \#2, infinite inbox~\cite{morehouse2025stalling,morehouse2025failback2,morehouse2025inbox} (2025-12-04) & LINE & 3 & 0.19.0 & 196 & 28 (12) & 22 (10) & 13 (6) \\
LND & Zero-timestamp gossip panic~\cite{lightninglabs2026zerots} (2026-06-18) & PARTIAL & 1 & 0.20.1 & 126 & 34 [34--97] (27) & 25 (16) & 20 (10) \\
LND & Lightning Labs batch: gossip, peer and queue DoS~\cite{lightninglabs2026aug} (2026-08-11) & LINE & 6 & 0.19.0 & 446 & 15 (6) & 13 (6) & 13 (6) \\
LND & HTLC first-stage sweep budget failure~\cite{lightninglabs2026sweepbudget} (2026-08-11) & LINE & 1 & 0.19.0 & 446 & 10 (5) & 8 (6) & 8 (6) \\
LND & Too few confirmations on cooperative close~\cite{teinturier2026closeconf} (2026-08-13) & LINE & 1 & 0.21.0 & 69 & 51 (53) & 34 (29) & 33 (28) \\
LND & Gossip stall via malformed announcements~\cite{lightninglabs2026gossipstall} (2026-09-21) & PARTIAL & 1 & 0.20.1 & 221 & 8 [8--32] (5) & -- & 8 (4) \\
LND & Brontide write-allocation OOM~\cite{lightninglabs2026brontide} (2026-09-21) & LINE & 1 & 0.20.0 & 305 & 22 (11) & -- & 20 (10) \\
\bottomrule
\end{tabular}}
\par\vspace{2pt}\parbox{\textwidth}{\footnotesize $^\dagger$Before July 2024 our view did not include the few live nodes on LND releases older than 0.14 (\S\ref{sec:data}), so these exposures are lower bounds and the later rise to 1\% is the wider view, not nodes moving onto affected releases. In 2023 the mixed LND line 0.15.1--0.17 held almost every LND node, so the bounds of the 2023 LND PARTIAL rows span 0--100\%. --: no usable snapshot within 7 days.}
\end{table*}

%% file: tables/table_response.tex
\begin{table*}[t]
\centering\footnotesize
\caption{Did operators respond to the disclosure? Weekly share of live nodes on a wholly affected line that moved to a later line, in the 8 weeks before and the 4 weeks after disclosure (fewer where snapshots are missing or the cutoff intervenes), and the ratio after/before with a quasi-Poisson 95\% CI. ``Outside platform releases'' leaves out moves signed within 48~h after an Umbrel or BTCPay release that ships the new line (Section~\ref{sec:waves}); $p$: exact one-sided permutation test that the weekly rates after exceed those before, outside platform releases. Indented: the nodes of the row above that were one line behind the fix (Section~\ref{sec:response}). Managed fleet excluded.}
\label{tab:response}
\resizebox{\textwidth}{!}{%
\begin{tabular}{llrrrrrllr}
\toprule
 & & Fix$\to$disc. & At risk & Weeks & \multicolumn{2}{c}{Weekly rate (\%)} & \multicolumn{3}{c}{Rate ratio after/before} \\
\cmidrule(lr){6-7}\cmidrule(lr){8-10}
Impl. & Advisory (disclosed; affected lines) & (days) & nodes & bef./aft. & before & after & all moves & outside platform releases & $p$ \\
\midrule
CLN & connectd queue memory exhaustion (2026-07-19; $<$25.09) & 320 & 186 & 7/4 & 1.6 & 5.4 & 3.45 [0.98, 12.08] & 1.54 [0.47, 5.09] & 0.212 \\
CLN & gossipd unknown-SCID memory exhaustion (2026-07-19; $<$26.04) & 90 & 520 & 7/4 & 1.6 & 6.2 & 3.87 [2.43, 6.17] & 3.28 [2.14, 5.04] & 0.003 \\
 & \quad of which one line behind (2026-07-19; 25.09--25.12) & 90 & 334 & 7/4 & 1.8 & 6.8 & 3.71 [2.39, 5.75] & 3.87 [2.33, 6.46] & 0.003 \\
CLN & Ping-flood OOM (2026-08-25; $<$25.09) & 357 & 125 & 8/4 & 3.4 & 5.0 & 1.48 [0.36, 6.13] & 1.99 [1.03, 3.84] & 0.077 \\
LND & Irrevocable fees / update\_fee breach fee-burn (2024-12-10; $<$0.18) & 89 & 1,861 & 8/3 & 1.2 & 1.4 & 1.24 [0.89, 1.73] & 1.26 [0.87, 1.84] & 0.133 \\
LND & gossip\_timestamp\_filter goroutine DoS (2025-07-22; $<$0.18) & 313 & 542 & 8/3 & 0.7 & 0.9 & 1.25 [0.73, 2.14] & 1.19 [0.71, 2.00] & 0.327 \\
LND & Replacement stalling, failback \#2, infinite inbox (2025-12-04; $<$0.19) & 196 & 1,273 & 8/4 & 1.6 & 1.6 & 0.97 [0.69, 1.37] & 0.86 [0.62, 1.19] & 0.810 \\
LND & Zero-timestamp gossip panic (2026-06-18; $<$0.20) & 126 & 1,420 & 8/4 & 1.3 & 1.6 & 1.30 [0.84, 2.01] & 1.15 [0.80, 1.66] & 0.214 \\
LND & Lightning Labs batch: gossip, peer and queue DoS (2026-08-11; $<$0.19) & 446 & 546 & 8/4 & 1.4 & 1.4 & 0.99 [0.22, 4.56] & 1.30 [0.56, 3.03] & 0.295 \\
LND & HTLC first-stage sweep budget failure (2026-08-11; 0.18) & 446 & 363 & 8/4 & 1.7 & 1.8 & 1.06 [0.21, 5.21] & 1.48 [0.58, 3.78] & 0.232 \\
LND & Too few confirmations on cooperative close (2026-08-13; $<$0.21) & 69 & 1,821 & 8/4 & 6.5 & 3.9 & 0.60 [0.20, 1.78] & 0.73 [0.37, 1.44] & 0.891 \\
LND & Gossip stall via malformed announcements (2026-09-21; 0.19) & 221 & 292 & 8/3 & 5.3 & 1.1 & 0.20 [0.01, 2.97] & 0.32 [0.11, 0.99] & 0.976 \\
LND & Brontide write-allocation OOM (2026-09-21; $<$0.20) & 305 & 763 & 8/3 & 3.2 & 0.8 & 0.25 [0.02, 2.92] & 0.41 [0.16, 1.09] & 0.970 \\
\bottomrule
\end{tabular}}
\end{table*}

%% file: tables/table_robustness.tex
\begin{table*}[t]
\centering\footnotesize
\caption{Robustness of the case-study exposures: share of live nodes on an affected line (\%) at disclosure and at the cutoff, with live capacity in parentheses for the main definition. Liveness: a peer-set channel direction toward the node enabled and updated within 7, 14 (main) or 28 days. Strict bounds: nodes whose every consistent line is affected, and nodes with any affected consistent line. Last column: share of the exposed nodes whose current announcement was signed in the previous 14 days, against 94--96\% of all live LND and 15--22\% of all live CLN nodes on the same days (Section~\ref{sec:robust}).}
\label{tab:robustness}
\resizebox{\textwidth}{!}{%
\begin{tabular}{lllrrrrrr}
\toprule
Advisory & When & Day & 14\,d (main) & 7\,d & 28\,d & Strict bounds & Fleet as nodes & Signed $<$14\,d \\
\midrule
LND $<$ 0.19, Morehouse, 4 Dec 2025 & disclosure & 2025-12-04 & 28 (12) & 27 & 28 & [7, 28] & 21 & 89 \\
 & cutoff & 2026-10-13 & 13 (6) & 12 & 13 & [4, 13] & 11 & 74 \\
LND $<$ 0.19, Lightning Labs, 11 Aug 2026 & disclosure & 2026-08-11 & 15 (6) & 14 & 15 & [5, 15] & 13 & 77 \\
 & cutoff & 2026-10-13 & 13 (6) & 12 & 13 & [4, 13] & 11 & 74 \\
CLN $<$ 25.09, connectd DoS, 19 Jul 2026 & disclosure & 2026-07-19 & 24 (20) & 24 & 24 & [24, 24] & -- & 12 \\
 & cutoff & 2026-10-13 & 12 (2) & 12 & 12 & [12, 12] & -- & 15 \\
CLN $<$ 26.04, gossipd DoS, 19 Jul 2026 & disclosure & 2026-07-19 & 67 (58) & 66 & 67 & [63, 100] & -- & 10 \\
 & cutoff & 2026-10-13 & 24 (13) & 24 & 24 & [21, 100] & -- & 14 \\
\bottomrule
\end{tabular}}
\end{table*}

%% file: sections/discussion.tex
\section{Discussion}
\label{sec:discussion}

\paragraph{Platforms, not advisories, move nodes}
The largest upgrade events in our data are platform pushes. Each time the Umbrel app store shipped a new
LND line, \WaveUmbrelMinNodes--\WaveUmbrelMaxNodes{} nodes re-announced within about a day; BTCPay Server's security
release moved \WaveBTCPayNodes{} LND nodes in a day, and with them the CLN nodes on old lines that CLN's own
disclosure had not moved. No LND disclosure produced a visible increase at all. Frei et al.\ found the same for browsers:
the update mechanism, more than the user, sets the patch level~\cite{frei2009firefox}; and Sarabi et al.\ found the
vendor's patch speed to matter more than users' promptness~\cite{sarabi2017patchme}. On the Lightning Network the
vendor that reaches most small nodes is the platform, not the implementation. The same division, between maintainers
who fix and disclose and packagers who deliver, runs through much open-source software, where distributions and app
stores stand between a project and its users; there too, the packager's schedule is likely to set the patch level.
Each Umbrel wave came
\WaveUmbrelDaysMin--\WaveUmbrelDaysMax{} days after LND's own release, a large part of the median time to upgrade
(\MedianLNDAll{} days). Two practical consequences follow. A fix reaches most small nodes when platforms ship it, so
the platforms' release schedule, not only the implementation's, decides how long a vulnerability stays open. And a
platform that tells its users to update moves even nodes that ignored earlier releases: \WaveBTCPayOlder{} of the
\WaveBTCPayNodes{} nodes in the BTCPay wave had skipped at least one LND line, although the advisory that moved them
concerned BTCPay rather than LND.

\paragraph{Disclosure reaches the operators who were already upgrading}
The LND advisories of 2025 and 2026 were published long after the fixes shipped (median \FixToDiscMedian{} days across
the LND and CLN advisories), a common practice in coordinated vulnerability disclosure that lets operators upgrade
before attackers learn what to look for. It works for
the operators who follow releases: half of LND nodes move to a new line within three months, and the largest move
first. The nodes still on an affected line at disclosure are the ones that did not, and for LND the disclosure did not
detectably change their pace: \PctCohortStillLLBatch\% of those exposed to Lightning Labs' August batch were still exposed two
months later. The one disclosure that did move operators makes the same point from the other side. CLN's July
disclosures raised the rate of leaving affected lines \RROpClnRecent-fold among nodes one line behind, operators who
keep up; the nodes further behind moved when BTCPay's advisory had their operators update. CLN's emergency call of
26~August selected the same way: the nodes that went offline were mostly on the newest line, already patched against
every vulnerability CLN had disclosed, while most nodes still exposed to the July vulnerabilities stayed online. Read
through release lines, the shutdown we measured earlier~\cite{kumar2026clientdiversity} was followed mainly by the
operators who were already keeping up.
Reported attacks do not
explain the CLN response, since the maintainers reported attacks on unpatched nodes only in
October~\cite{coreln2026attacks}. More plausibly, a fix that had shipped \FixToDiscClnGossipd{} days before disclosure
found most exposed CLN nodes one line behind, while one shipped \FixToDiscLLBatch{} days before found only the LND
nodes that had skipped every release since. Notification studies on the Internet find that most notified operators
never act~\cite{li2016notifications} and that direct notification raised Heartbleed patching by about
half~\cite{durumeric2014heartbleed}. A Lightning advisory is not a notification: it is a post that reaches those who
read the implementation's channels, and the operators who do not are reached, if at all, by their platform.
Release-line inference also gives maintainers a measurement they have lacked: the share of live nodes and capacity
still on an affected line, read in aggregate from public gossip, which can inform when to disclose.

\paragraph{The tail}
What remains exposed are small nodes, nodes that skipped earlier releases, and nodes that no platform updates. For
advisories that require a channel with the victim, these nodes are also the cheapest to reach. Their share of
capacity at risk is small (\ExpCapLLBatchDisc\% of live LND capacity for the August~2026 LND batch), but each holds its
operator's funds.

\paragraph{Counting nodes}
At its peak, one coordinated fleet was more than a quarter of the live LND nodes we observe, with almost no
capacity. Any statistic that counts nodes, whether implementation shares, version shares or exposure, would have been
dominated by that fleet's schedule. Fleets are cheap to create and hard to see. This one gave itself away by a shared
alias and hubs~\cite{kumar2026clientdiversity} and, here, by upgrading in step; a fleet with neither would pass for
hundreds of independent operators. Node-level Lightning measurements should look for such fleets before counting,
and report capacity alongside nodes.

\section{Limitations}
\label{sec:limitations}
\emph{Release lines, not patches.} Gossip cannot separate patch releases, so fixes inside a line, including all
of CLN's 2026 patch releases, cannot be measured; for advisories with a fix inside a line we report bounds.
\emph{Inference.} An operator can switch features on or off so that an older line looks like a newer one or the
reverse; the strict range covers every non-developer configuration, and validation on self-identifying CLN nodes
bounds the error of the likely line. For LND we have no ground truth at scale, only the three checks of
\S\ref{sec:validation}. A node that upgrades without re-announcing, or whose announcement we do not receive, keeps its
old bits, which biases exposure up.
\emph{Platforms.} We can date the releases of two platforms; upgrades pushed by others count as operators' own, so
platform shares are lower bounds. Dropping every move in the 48~hours after a platform release also drops operators
who moved on their own in those hours, which makes the ratios outside platform releases conservative.
\emph{Response design.} The before/after comparison is biased towards ratios below one when a line is young. With
\NResp{} disclosures, a few weeks on each side and moves that cluster, the intervals are wide: a ratio near one rules
out a large response, not a small one.
\emph{Vantage and liveness.} All data come from one node's view of the public graph; unannounced nodes, including
most mobile wallets, are not covered, and liveness is inferred from peers' channel updates.
\emph{Fleets.} We identified one managed fleet by its alias and hubs. Smaller fleets without a shared alias would go
undetected and be counted as independent nodes.
\emph{Exposure is not exploitability.} Several advisories require a channel with the victim, a particular
configuration or inbound connections; we count nodes on affected releases, not nodes that can be attacked. Our
advisory list is compiled from public sources and may miss silent fixes.
\emph{Association.} The hazard model describes which nodes upgrade sooner; it does not identify why.

\section{Ethics}
\label{sec:ethics}
Release-line inference identifies nodes running vulnerable software, and in 2026 Lightning deployments were
exploited (BTCPay Server~\cite{btcpay2026advisory}) and CLN maintainers reported attacks on unpatched
nodes~\cite{coreln2026attacks}. We therefore publish aggregates only: no list, ranking, filter or label of individual
nodes by release, platform or exposure, in the paper or in the released data. All inputs are public gossip and public
release records, every advisory we study had been published before our analysis began, and the paper reports no new
vulnerability.

\section{Conclusion}
\label{sec:conclusion}
Lightning gossip carries no version string, but the feature bits each release announces identify its release line
closely enough to measure who patches the network. Operators patch on release schedules: most upgrade within a few
months of a release. Platforms patch on theirs: the largest single movements are node platforms shipping a release
to all their users at once. Disclosures, which typically come long after the fix, move few nodes: of those we
can test, only Core Lightning's of July~2026 raised the pace, and only among operators who were already close to
current. A tail of small and neglected nodes stays on vulnerable releases. Measured by capacity it is small; measured
by nodes, it depends on whether one counts a single fleet as hundreds of operators. For the maintainers of Lightning
implementations, the lever that moves the tail is the platforms that package their releases.